\documentclass[sn-mathphys-num]{sn-jnl}

\usepackage{graphicx}%
\usepackage{multirow}%
\usepackage{amsmath,amssymb,amsfonts}%
\usepackage{amsthm}%
\usepackage{mathrsfs}%
\usepackage[title]{appendix}%
\usepackage{xcolor}%
\usepackage{textcomp}%
\usepackage{manyfoot}%
\usepackage{booktabs}%
\usepackage{algorithm}%
\usepackage{algorithmicx}%
\usepackage{algpseudocode}%
\usepackage{listings}%
\usepackage{utfsym}
\usepackage{geometry}
\theoremstyle{thmstyleone}%
\theoremstyle{thmstyletwo}%

\theoremstyle{thmstylethree}%

\begin{document}

\title[Article Title]{CPE-Pro: A Structure-Sensitive Deep Learning Method for Protein Representation and Origin Evaluation}
\author[1]{\fnm{Wenrui} \sur{Gou}}\email{gwr@mail.ecust.edu.cn}
\equalcont{These authors contributed equally to this work.}
\author[1]{\fnm{Wenhui} \sur{Ge}}\email{gwh@mail.ecust.edu.cn}
\equalcont{These authors contributed equally to this work.}
\author[1]{\fnm{Yang} \sur{Tan}}\email{tyang@mail.ecust.edu.cn}
\author[1]{\fnm{Mingchen} \sur{Li}}\email{lmc@mail.ecust.edu.cn}
\author*[1]{\fnm{Guisheng} \sur{Fan}}\email{gsfan@ecust.edu.cn}
\author*[1]{\fnm{Huiqun} \sur{Yu}}\email{yhq@ecust.edu.cn}

\affil*[1]{\orgdiv{School of Information Science and Engineerin}, \orgname{East China University of Science and Technology}, \orgaddress{\street{Street}, \city{Shanghai}, \postcode{200237}, \country{China}}}

\abstract{Protein structures are important for understanding their functions and interactions. Currently, many protein structure prediction methods are enriching the structure database. Discriminating the origin of structures is crucial for distinguishing between experimentally resolved and computationally predicted structures, evaluating the reliability of prediction methods, and guiding downstream biological studies. Building on works in structure prediction, We developed a structure-sensitive supervised deep learning model, \textbf{C}rystal vs \textbf{P}redicted \textbf{E}valuator for \textbf{Pro}tein Structure (\textbf{CPE-Pro}), to represent and discriminate the origin of protein structures. CPE-Pro learns the structural information of proteins and captures inter-structural differences to achieve accurate traceability on four data classes, and is expected to be extended to more. Simultaneously, we utilized Foldseek to encode protein structures into ``structure-sequence'' and trained a protein \textbf{S}tructural \textbf{S}equence \textbf{L}anguage \textbf{M}odel, \textbf{SSLM}. Preliminary experiments demonstrated that, compared to large-scale protein language models pre-trained on vast amounts of amino acid sequences, the ``structure-sequence'' enables the language model to learn more informative protein features, enhancing and optimizing structural representations. We have provided the code, model weights, and all related materials on \url{https://github.com/GouWenrui/CPE-Pro-main.git}.}


\maketitle

\section{Introduction}\label{sec1}

The function of a protein is determined by its folded structure; therefore, characterizing protein structure is crucial for understanding and studying its biological functions. Protein folding is a complex and highly coordinated process that determines the active sites, ligand-binding capabilities, and functional roles of the protein both inside and outside the cell.\cite{gold2006fold} Even slight alterations in structure can lead to significant changes in protein function, potentially resulting in disease. Consequently, a comprehensive understanding of protein folding is not only essential for elucidating its biological functions but also serves as a guide for research and applications in drug design, disease diagnosis, and treatment.

However, experimentally determining protein structures poses numerous challenges. Despite significant efforts by scientists over the past few decades, the number of experimentally determined structures reported in the Protein Data Bank (PDB)\cite{bernstein1977protein} remains far lower than the number of known protein sequences, creating a substantial data gap. The limited availability of protein structure data significantly hinders a comprehensive understanding of protein functions and interaction mechanisms. To address this limitation, researchers have increasingly turned to computational methods, such as protein folding simulations (AMBER\cite{case2005amber}, GROMACS\cite{abraham2015gromacs}), to guide the determination of protein structures. One of the most representative methods is homology modeling\cite{altschul1990basic,waterhouse2018swiss}. However, homology modeling relies on the similarity to known structures and cannot be accurately applied to proteins without known similar structures. While these computational methods somewhat alleviate the issue of insufficient experimental data, they still do not fully resolve the broader applicability of structure prediction, particularly for novel or complex proteins, where significant limitations remain.

As artificial intelligence technology advances, an increasing number of supervised\cite{luo2021ecnet,li2023sesnet} and unsupervised\cite{rives2021biological,meier2021language, lin2023evolutionary} deep learning models are showing promise in the field of protein research, significantly enhancing the efficiency and accuracy of structure prediction. However, despite the proliferation of structure prediction models\cite{yang2020improved,jumper2021highly,lin2022language}, a comprehensive evaluation of the differences and distinguishability between predicted and experimentally determined structures has yet to be undertaken. Significant discrepancies may still exist between prediction results and experimental findings, arising from the limitations of prediction models in handling the complex dynamics and interactions involved in protein folding. For example, models may fail to capture subtle changes in the protein folding process or perform poorly when dealing with proteins that have not been extensively studied. Ignoring these differences could lead to misunderstandings of protein function and behavior, potentially negatively impacting subsequent biological applications. Additionally, malicious injection in structural databases may further exacerbate inaccuracies in model downstream performance, leading to a decrease in the reliability of prediction results. Therefore, it is crucial to systematically evaluate the differences between predicted and experimental structures and to develop effective methods to distinguish the origin of protein structures. Our research aims to address this gap and propose a general and reliable solution.

Our contributions are as follows:
\begin{itemize}
    \item We introduce CPE-Pro, a model that excels in distinguishing between crystal and predicted protein structures by learning structural features.

    \item Using the CATH 4.3.0\cite{orengo1997cath} non-redundant dataset, we create the protein folding dataset CATH-PFD with multiple prediction models.

    \item Preliminary experiments indicate that, compared to amino acid sequences, ``structure-sequences'' enable language models to learn more effective protein feature information, enriching and optimizing structural representations when integrated with graph embeddings.

    \item We have open-sourced the code, model weights, and CATH-PFD dataset for CPE-Pro, providing valuable resources for protein structure research.
\end{itemize}

\section{Related Works}\label{sec2}

\subsection{Protein Representation Learning}\label{subsec2-1}

\begin{figure}[h]
\centering
\includegraphics[width=0.9\textwidth]{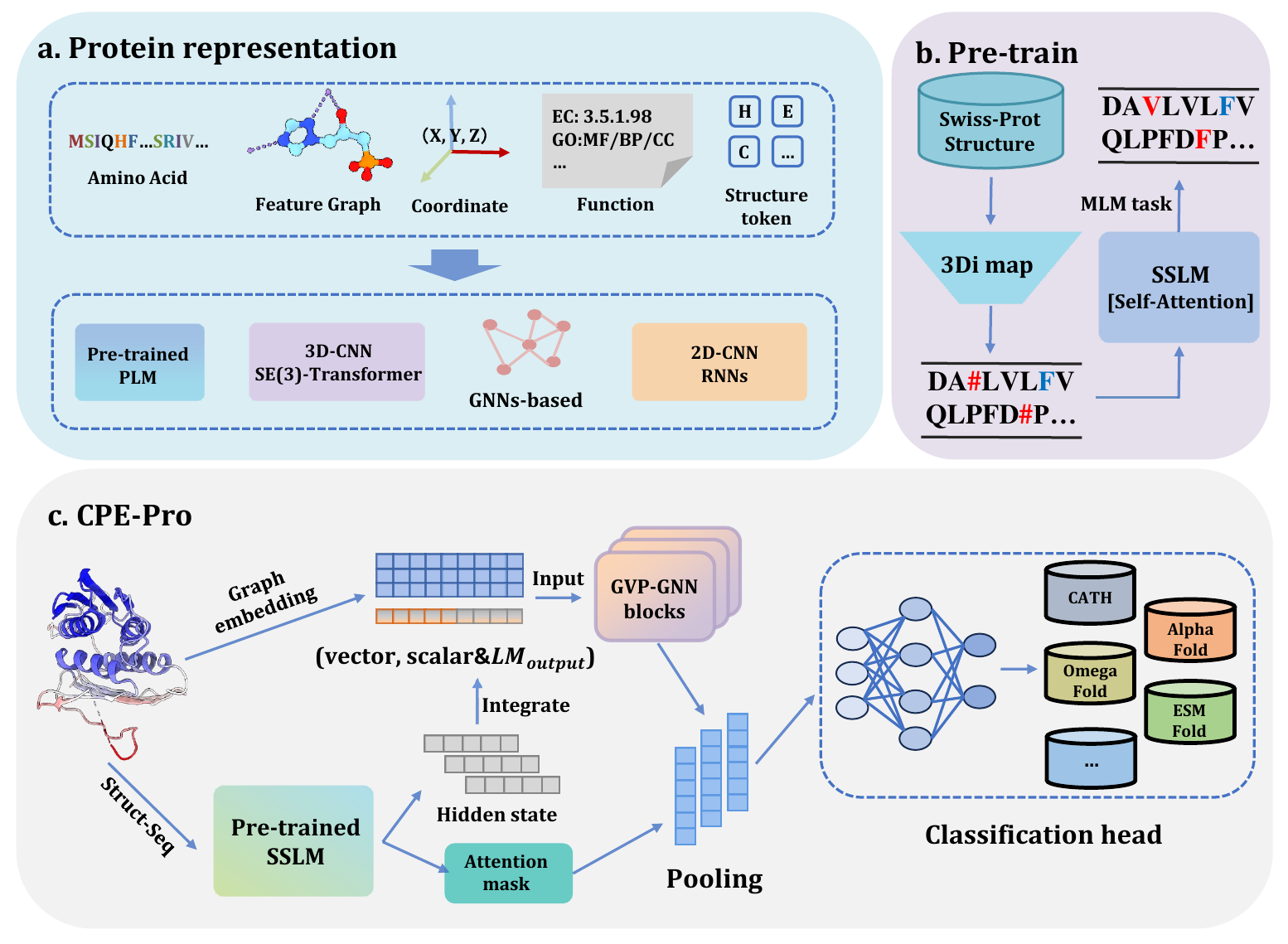}
\caption{\textbf{a. Protein representation methods.} Proteins can be input into the model in various forms, including amino acid sequences, feature maps, three-dimensional coordinates, functional descriptions, and sequences composed of structural tokens, capturing the multi-level features of proteins. \textbf{b. Pre-training of SSLM.} SSLM is pre-trained on over 100,000 protein structures from the Swiss-Prot database and trained on various masked language modeling tasks, learning the relationships between ``structure-sequences'' and their corresponding three-dimensional structural features, thereby effectively representing protein structural information. \textbf{c. CPE-Pro model architecture.} The CPE-Pro model integrates a pre-trained protein structure language model with a graph embedding module, inputting the combined representation into the GVP-GNN module for computation. The pooling module aggregates structural information using attention masking, enhancing the quality of the representation. Ultimately, a multilayer perceptron serves as the source discriminator, outputting predicted probabilities.}\label{model-architecture}
\end{figure}

In various biological tasks, it is essential to learn effective protein representations, such as for predicting protein functions or the effects of mutations. Protein representation methods can be classified into three approaches based on different modalities: sequence-based, structure-based, and a combination of sequence and structure.Figure \ref{model-architecture}.a shows various protein representation methods.

\textbf{Sequence-based.} With the rise of deep learning and advances in high-throughput sequencing technologies, data-driven methods are gradually replacing traditional analyses based on biological or physical priors. Protein sequences can be viewed as a form of ``biological text'', and convolutional neural network\cite{lecun1998gradient} can directly capture local dependencies between amino acids. Techniques from natural language processing are also widely applied in protein representation learning. Models for individual protein sequence modeling include Variational Auto-Encoders (VAE)\cite{kingma2013auto}, long short-term memory networks (LSTM)\cite{graves2012long}, and large pre-trained protein language models(PLMs) like ESM-1b\cite{rives2021biological} based on the Transformer architecture\cite{vaswani2017attention}. Some studies have employed GPT\cite{radford2018improving}-based architectures for sequence modeling, using generative pre-training to represent protein sequences and make predictions, such as ProtGen2\cite{nijkamp2023progen2} and ProGPT2\cite{ferruz2022protgpt2}. Compared to single sequences, multiple sequence alignment (MSA)\cite{katoh2002mafft} input aims to capture co-evolutionary information from a set of evolutionarily related sequences. In this context, the MSA Transformer\cite{rao2021msa} utilizes row and column attention mechanisms to model a set of protein sequences, allowing it to simultaneously consider inter-sequence relationships and conservation. Additionally, some research integrates protein sequences with other types of information, such as converting gene ontology annotations into fixed-size binary vectors for joint input with sequences in ProteinBERT\cite{brandes2022proteinbert}, enabling the model to leverage the connections between sequence and functional annotations effectively.

\textbf{Structure-based.} While sequence-based research methods have been shown to be effective in several studies, the structural information of proteins is a critical determinant of their function. Models based on graph neural networks (GNNs) demonstrate significant advantages and broad applicability in representing protein three-dimensional structures. For example, ProteinGCN\cite{sanyal2020proteingcn} constructs spatially adjacent protein graphs and is trained on tasks related to both local and global accuracy in protein modeling. Pre-trained protein representation methods\cite{zhang2022protein} based on three-dimensional structures represent proteins as residue graphs, where nodes correspond to the three-dimensional coordinates of \( \alpha \) carbons and edges represent relationships between residues. In modeling protein secondary structures, these are often transformed into secondary structure sequences, with \( \alpha \)-helices, \( \beta \)-sheets, and random coils represented as token sequences. DeepSS2GO\cite{song2024deepss2go} utilizes one-hot matrices to model secondary structures for predicting key protein functions. SES-Adapter\cite{tan2024simple} enhances the performance of PLMs in downstream tasks by integrating secondary structure sequence embeddings with other types of embeddings through a cross-attention mechanism.

\textbf{Integrating Sequence and Structure.} Combining protein sequence and structural information not only considers the sequential characteristics of amino acids but also reveals their spatial interactions and arrangements. In this regard, DeeFRI\cite{gligorijevic2021structure} combines LSTM and graph convolutional networks (GCNs)\cite{kipf2016semi} to jointly learn complex structure-function relationships. LM-GVP\cite{wang2021lm} modifies the input variables of GVP\cite{jing2020learning} by using sequence embeddings generated by ProteinBERT as input. ESM-GearNet\cite{zhang2023systematic} designs various fusion methods between PLMs and GearNet to investigate the effectiveness of different fusion strategies. ProtSSN\cite{tan2023semantical} combines ESM2\cite{lin2023evolutionary} with equivariant graph neural networks (E-GNNs)\cite{satorras2021n} to extract geometric features of proteins, aiming to accurately predict biological activity and thermal stability. ProstT5\cite{heinzinger2023prostt5} introduces 3Di (Three-Dimensional Indexing) alphabet used by Foldseek\cite{van2024fast} based on the ProtT5-XL-U50 model\cite{elnaggar2021prottrans} and is trained on labeling translation tasks in two modes. Additionally, SaProt\cite{su2023saprot} creates a larger structural-aware vocabulary using two types of labels and is pre-trained on a large-scale protein dataset for masked language tasks. Compared to previous work, the SSLM in CPE-Pro uses only the ``structure-sequence'' builts from 3Di alphabet for pre-training on Swiss-Prot, without relying on amino acid labels, and effectively learns protein representations by integrating existing structural information with GVP.

\subsection{Protein Structure Prediction}\label{subsec2-2}

Amino acids form the linear sequence of proteins, but they acquire activity and biological function only when folded into specific spatial conformations. Early structure prediction methodsolded into specific spatial conformations. Early structure prediction methods\cite{sievers2014clustal, riesselman2018deep} primarily relied on sequence similarity, utilizing the homology of known protein sequences for predictions. These methods infer the structure of the target protein by aligning the sequence of interest with known homologous sequences and using the structural information from these homologs. Compared to sequence-based methods, structure-based learning approaches theoretically offer better solutions for acquiring protein information. In recent years, advancements in deep learning technologies have led to breakthrough progress in protein structure prediction.


In the study of secondary structure prediction, DeepCNF\cite{wang2016protein} integrates conditional random fields (CRFs)\cite{lafferty2001conditional} and shallow neural networks to successfully model the interdependencies between adjacent secondary structure labels. SSREDNs\cite{wang2017protein} leverage deep and recurrent structures to simulate the complex nonlinear mapping relationships between input protein features and secondary structures, while also capturing interactions among consecutive residues in the protein chain.


More research has focused on three-dimensional structure prediction. In this context, trRosetta \cite{yang2020improved} utilizes co-evolution data, combined with deep residual networks, to predict the orientation and distances between residues through Rosetta-constrained energy minimization protocols. trRosettaX-single\cite{wang2022single} focuses on the structure prediction of single-chain proteins. Regarded as a "milestone," AlphaFold2 \cite{jumper2021highly} significantly enhances the accuracy and breadth of protein structure prediction by embedding multiple sequence alignments and paired features through Evoformer. OmegaFold \cite{wu2022high} employs GCNs and self-attention \cite{vaswani2017attention} to effectively capture both global and local features of protein sequences, excelling in handling proteins of varying lengths and complexities. ESMFold \cite{lin2022language} adopts a large-scale Transformer architecture, trained on extensive protein sequence datasets, to extract deep evolutionary features from sequence information without relying on multiple sequence alignments, demonstrating exceptional performance in predicting new structural domains and distant homologs. Leveraging the predictions from these models will provide robust support for biomedical research, drug development, and further advancements across various scientific domains.

\section{Materials \& Methods}\label{sec3}

\subsection{Dataset}\label{subsec3-1}


In this study, we utilized the non-redundant dataset from version 4.3.0 of the CATH (Class, Architecture, Topology, Homologous superfamily) database as our benchmark dataset. CATH is a significant protein structure classification database that categorizes proteins based on their structural and functional features through a systematic hierarchical structure. The high-resolution three-dimensional structure data provided by these experimental techniques are stored in the Protein Data Bank (PDB). The non-redundant dataset has a sequence identity filtering threshold of 40\%, ensuring no high sequence similarity between structural domains. By removing redundant protein structures, each entry in the database represents a unique structural representation. 

We extracted the amino acid sequences of proteins from the benchmark dataset. Using multiple state-of-the-art protein structure prediction models, we predicted the structures corresponding to these amino acid sequences. These structures were organized and categorized based on individual proteins and prediction models to construct a Protein Folding Dataset, CATH-PFD, which will be used for training and validating CPE-Pro. Table \ref{tab1} presents detailed information about the dataset CATH-PFD.

\begin{table}[h]
\caption{The information on CATH-PFD}\label{tab1}%
\begin{tabular}{lccc}
\toprule
Model/Origin & Number & pLDDT(\%)  & Method to Obtain \\
\midrule
CATH     & 31,885   & \textbf{--} & Download from \url{https://www.cathdb.info/} \\
AlphaFold2   & 31,885 & 92.4 & Predicted using the ColabFold\cite{mirdita2022colabfold} \\
OmegaFold    & 31,881 & 82.7 & Predicted using the OmegaFold \\
ESMFold    & 23,912 & 76.2 & Predicted using the ESMFold \\
\botrule
\end{tabular}
\footnotetext{The number of structures from different origins, their pLDDT scores, and acquisition methods. The CATH database contains 31,885 protein structures. For structure prediction, we deployed OmegaFold, ESMFold, and the efficient AlphaFold2 variant, ColabFold, using publicly available resources from previous studies. Unpredictable or erroneous protein structure results were excluded during the prediction process, leading to differences in the number of entries across various categories.}
\end{table}

\subsection{Model Architecture}\label{subsec3-2}

Crystal vs Predicted Evaluator for Protein Structure (CPE-Pro), designed for discriminating structural origins, integrates two distinct structure encoders corresponding to graphical and sequential representations of the structures. Figure \ref{model-architecture}.c illustrates the detailed architecture of CPE-Pro.

CPE-Pro implements Geometric Vector Perceptrons – Graph Neural Networks (GVP-GNN)\cite{hsu2022learning}, which learn dual relationships and geometric representations of three-dimensional macromolecular structures as part of the protein structural encoder.

In obtaining information about the three-dimensional structure of proteins, we focus on a specific chain, ignoring the parts we are not concerned with. We concentrate on the coordinates of key atoms that constitute the protein backbone—N, CA, and C atoms—since these atoms are necessary for understanding the protein structure. Subsequently, we extract the geometric information of nodes and edges from the raw data and compute features such as distances and directions between them.For a protein graph \( G = (V, E) \), where \( V \) represents the set of nodes and \( E \) represents the set of edges, each node \( v \in V \) represents a residue of the protein, and each edge \( e \in E \) represents an interaction or spatial proximity relationship between residues. For the ith residue, the feature consists of scalars and vectors, i.e., \( \mathcal{R}_v^i = (r_s^i, r_{vec}^i) \). A embedding layer computes embeddings for the protein feature \( \mathcal{R}_v \), i.e.,

\begin{equation}
\mathcal{R}_v^{\prime} = GVP(r_s, r_{vec}).\label{eq1}
\end{equation}

\noindent The GVP (·) layer mainly carries out scalar-vector propagations, i.e.,

\begin{equation}
r_s^{\prime} = \sigma_1 \left(W_1 \cdot Concat\left(r_s, Norm\left(W_2 \cdot r_{vec}\right)\right) + b_1 \right).\label{eq2}
\end{equation}

\begin{equation}
r_{vec}^{\prime} = W_3 \cdot \left(W_2 \cdot r_{vec}\right) \odot \sigma_2 \left(W_4 \cdot r_s^{\prime} + b_2 \right).\label{eq3}
\end{equation}

\noindent Here, \(W_1\), \(W_2\), \(W_3\), \(W_4\), \(b_1\), \(b_2\) are learnable parameters for this layer, \(\sigma_1(\cdot)\) and \(\sigma_2(\cdot)\) denote activation functions. In each graph propagation step, messages from neighboring nodes and edges in the protein graph update the embedding of the current node,

\begin{equation}
r_{msg}^{j \rightarrow i} := GVP_{conv}\left(Concat\left( r_{v}^{j}, r_{e}^{j \rightarrow i}\right)\right).\label{eq4}
\end{equation}

\begin{equation}
r_{v}^{i} \leftarrow LayerNorm\left(r_{v}^{i} + \frac{1}{k} Drop\left( \sum_{\substack{j: e_{ji} \in E}} r_{msg}^{j \rightarrow i}\right)\right).\label{eq5}
\end{equation}

\noindent Here, \( r_{v}^{i} \) and \( r_{e}^{j \rightarrow i} \) are the embeddings of the node i and \( edge_{j \rightarrow i} \) as above, and \( r_{msg}^{j \rightarrow i} \) represents the message passed from node j to node i. \( GVP_{conv} \) denotes the stacked 3-layer GVP. we also add a feed-forward layer to update the node embeddings at all nodes i:

\begin{equation}
r_{v}^{i} \leftarrow LayerNorm\left(r_{v}^{i} + Drop\left(GVP_{conv}^{\prime}\left(r_{v}^{i}\right)\right)\right).\label{eq6}
\end{equation}

\noindent The \( GVP_{conv}^{\prime} \) is a stacked 2-layer \( GVP \). The GVP-GNN block is formed by stacking \( GVP \) convolution and feed-forward transformations, as defined in Eq.~(\ref{eq4})--(\ref{eq6}). To enhance node representations, this block is iterated \( T \) times, with \( T = 3\) specified for our experiments.


The other part of CPE-Pro's structural encoder is \textbf{S}tructural \textbf{S}equence \textbf{L}anguage \textbf{M}odel, \textbf{SSLM}. First, the efficient protein structure data search tool Foldseek is used to convert protein structures into ``structure-sequences''. The primary process involves mapping the amino acid backbone of the protein to the 3Di alphabet to achieve structural discretization. This reflects the tertiary interactions between amino acids and describes the geometric conformations of residues and their spatial neighbors.

Next, using the 3Di alphabet as the vocabulary of structural elements and based on the Transformer architecture, we pre-train a protein structural language model, SSLM, from scratch. This aims to effectively model ``structure-sequences'' of proteins. The pre-training process employs the classic masked language modeling (MLM) objective\cite{devlin2018bert}, predicting masked elements based on the context of the ``structure-sequence''. The probability distribution for predicting a masked element \( P(s_i \mid s_1, \dots, s_{i-1}, s_{i+1}, \dots, s_n) \) is used, where \( s_i \) is the masked structural element and \( s_i \),\dots,\( s_{i-1} \),\( s_{i+1} \),\dots,\( s_i \) are its contexts. The loss function is defined as follows:

\begin{equation}
\mathcal{L}_{\mathcal{MLM}} = -\sum_{i=1}^{x} y_i \log(\hat{y}_i).\label{eq7}
\end{equation}

\noindent where \( \hat{y}_i \) denotes the model's predicted label, indicating the probability that the i-th token in the structural element vocabulary is the masked element, and \( y_i \) denotes the true label. The loss is computed only on \( x \) elements that are masked.

\noindent \textbf{Encoder of CPE-Pro.} We integrated the ``structure-sequence'' representation output by the pre-trained protein structural sequence language model with the embedding of protein graph. The SSLM can learn the sequential relationships and proximity interactions between local structural elements from the ``structure-sequence''. When combined with three-dimensional topological information, this approach aims to enrich and optimize the representation of protein structures. Specifically, we combined the representations obtained from the SSLM and the graph embeddings obtained from the GVP embedding layer as follows:\( \mathcal{R}_{S_{seq}} \in R^{N_{S_{seq}} \times D_{S_{seq}}}\)represents the ``structure-sequence'' representation, where \( N_{S_{seq}} \) is the length of the ``structure-sequence'', and \( D_{S_{seq}} \) is the feature dimension of each element in the ``structure-sequence''. We align the representation dimensions of \( \mathcal{R}_{S_{seq}} \) with \( r_s \in R^{N \times D_{scalar}} \) using a linear transformation layer. Afterwards, \( \mathcal{R}_{S_{seq}} \) and \( r_s \) are fused with adaptive weights, i.e.,

\begin{equation}
\mathcal{R}_{Pro} = \mathcal{W} \cdot r_s + (1 - \mathcal{W}) \cdot \mathcal{R}_{S_{seq}}.\label{eq8}
\end{equation}

\noindent Here, \( \mathcal{W} \) is a learnable parameter. Finally, the topological structure and node features of graph data incorporating information from the ``structure-sequence'' are utilized to learn structural representations in GVP-GNN blocks. The aim is to enhance the accuracy of structural discrimination through enriched and deepened data representations.

\noindent \textbf{Representation Classification}. The protein structures are processed by the structural encoder of CPE-Pro to obtain feature vectors \( \mathcal{R}_{Pro}^{\prime} \). We designed a straightforward classification head to perform the final discrimination task. The classification head consists of three components: a pooling layer with attention\cite{tan2024protsolm}, multilayer perceptron (MLP), and output (activation) layer. It aims to simplify feature dimensions and enable efficient classification. The specific process is represented by Eq.~(\ref{eq9})., where a layer of MLP applies weight matrix \( W_5 \), bias term \( b_3 \), and \( ReLU \) function to transform inputs. The output of the MLP is then passed through an activation function (ACT) that utilizes both \( Sigmoid \) and \( Softmax \) functions. \( Sigmoid \) corresponds to the binary classification task (\textbf{C}rystal - \textbf{A}lphaFold2, \textbf{C-A}), while \( Softmax \) is used for the multi-class classification problem (\textbf{C}rystal - \textbf{M}ultiple prediction models, \textbf{C-M}). The \( Pred \) is the final output of CPE-Pro.

\begin{equation}
Pred = ACT\left( \ldots ReLU \left( W_5 \cdot Attention\left(\frac{1}{n} \sum_{i=1}^{n} \left(\mathcal{R}_{Pro}^{\prime}\right)^{i}\right) + b_3\right)\right)
.\label{eq9}
\end{equation}

\section{Experimental Setups}\label{sec4}

\subsection{Baseline Methods}\label{subsec4-1}

We compared CPE-Pro to various embedded-based deep learning methods. Our analysis includes pre-trained PLMs, such as ESM1b, ESM1v\cite{meier2021language}, ESM2, ProtBert\cite{elnaggar2021prottrans}, Ankh\cite{elnaggar2023ankh}, combined with GVP-GNN as a model with amino acid sequence and structure input, and SaProt.

\subsection{Training Setups}\label{subsec4-2}

We employed the AdamW\cite{loshchilov2017decoupled} optimizer, setting the learning rate between 0.0001 and 0.00005, depending on the task and dataset size. Additionally, we applied CosineAnnealingLR for learning rate decay to improve convergence. We applied a dropout rate of 0.2 at the output layer and the number of epochs was set between 50 and 100. The loss functions employed were binary cross-entropy (Eq.~(\ref{eq10})) and categorical cross-entropy (Eq.~(\ref{eq11})). We implemented early stopping based on validation metrics such as accuracy to prevent overfitting. All protein folding, pre-training, and experiments were conducted on 8 NVIDIA RTX 3090 GPUs. Table \ref{tab2} shows the sampling and partitioning of the discriminative task on CATH-PFD.

\begin{table}[h]
\caption{Dataset partitioning for discrimination tasks}\label{tab2}%
\begin{tabular}{lcccc}
\toprule
Task & Train terms & Valid terms & Test terms & Total \\ 
\midrule
C-A & 10,000(5,000×2) & 1,000(500×2) & 1,000(500×2) & 12,000 \\
C-M & 20,000(5,000×4) & 2,000(500×4) & 2,000(500×4) & 24,000 \\
\botrule
\end{tabular}
\footnotetext{The distribution of training, validation, and test terms for two classification tasks: C-A and C-M. Each task is divided into subsets for model development and evaluation, detailing the number of terms and the class composition within each subset.}
\end{table}

\begin{equation}
\mathcal{L}_{C-A} = -\left[ y\log(\hat{y}) + \left(1 - y\right)\log\left(1 - \hat{y}\right) \right]
.\label{eq10}
\end{equation}

\begin{equation}
\mathcal{L}_{C-M} = -\sum_{i=1}^{n} y_i\log\left(\hat{y}_i\right).\label{eq11}
\end{equation}

\subsection{Evaluation Metrics}\label{subsec4-3}

Performance of pre-training tasks on SSLM is measured using perplexity\cite{brown1990statistical}, an indicator of a language model's predictive capability for a given text sequence. It quantifies the uncertainty of the model's probability predictions for each token. Let SSLM assign a probability \( P(S) \) to a ``structure-sequence'' $\langle s_1, \dots, s_{i-1}, s_{i+1}, \dots, s_n \rangle$ of length n. The perplexity of the model for the ``structure-sequence'' is defined as:

\begin{equation}
PP(S) = P^{-\frac{1}{N}}(S) = \left(\prod_{i=1}^{n} P\left(s_i \mid s_1, \dots, s_{i-1}, s_{i+1}, \dots, s_n\right)\right)^{-\frac{1}{N}}
.\label{eq12}
\end{equation}

The experiment reported several metrics to evaluate the performance of different models, including accuracy (ACC), Precision, Recall, F1-Score (F1), and Matthews correlation coefficient (MCC). Their calculation equations are as follows:

\begin{equation}
ACC = \frac{TP + TN}{TP + FP + FN + TN}
.\label{eq13}
\end{equation}

\begin{equation}
Precision = \frac{TP}{TP + FP}
.\label{eq14}
\end{equation}

\begin{equation}
Recall = \frac{TP}{TP + FN}
.\label{eq15}
\end{equation}

\begin{equation}
F1 = \frac{2 \times TP}{2 \times TP + FP + FN}
.\label{eq16}
\end{equation}

\begin{equation}
MCC = \frac{TP \times TN - FP \times FN}{\sqrt{(TP + FN) \times (TP + FP) \times (TN + FN) \times (TN + FP)}}
.\label{eq17}
\end{equation}

The terms TP, TN, FP, and FN denote the counts of correctly predicted positives, correctly predicted negatives, incorrectly predicted positives, and incorrectly predicted negatives, respectively.

\definecolor{third_score}{RGB}{128, 0, 128}
\begin{sidewaystable}
\caption{Performance Comparison of Baseline Models on CATH-PFD for Structure Discrimination}\label{main_result}
\begin{tabular}{lclccccccccccccc}
\toprule
\multicolumn{3}{c}{Method Information} & \multicolumn{3}{c}{Input} & \multicolumn{2}{c}{Acc(\%) $ \uparrow $
} & \multicolumn{2}{c}{Precision(\%) $ \uparrow $
} & \multicolumn{2}{c}{Recall(\%) $ \uparrow $
} &\multicolumn{2}{c}{F1-score $ \uparrow $
} & \multicolumn{2}{c}{MCC $ \uparrow $
} \\
\cmidrule(lr){1-3} \cmidrule(lr){4-6} \cmidrule(lr){7-8} \cmidrule(lr){9-10} \cmidrule(lr){11-12}  \cmidrule(lr){13-14}\cmidrule(lr){15-16}
Model & LM Version & Param & AA & Structure & Struct-seq & C-A & C-M & C-A & C-M & C-A & C-M & C-A & C-M & C-A & C-M\\
\midrule
GVP-GNN w/ & & & & & & & & & & & & & & & \\
ESM1b & t33 & 655M & \usym{2713} & \usym{2713} & \usym{2717} & \textbf{\textcolor{third_score}{93.9}} & \textbf{\textcolor{red}{92.0}} & 90.7 & \textbf{\textcolor{third_score}{92.0}} & 97.8 & \textbf{\textcolor{red}{92.0}} & \textbf{\textcolor{third_score}{0.941}} & \textbf{\textcolor{red}{0.920}} & \textbf{\textcolor{third_score}{0.881}} & \textbf{\textcolor{third_score}{0.893}} \\
ESM1v & t33 & 655M & \usym{2713} & \usym{2713} & \usym{2717} & 92.5 & 90.0 & 88.3 & 90.2 & 98.0 & 90.0 & 0.929 & 0.899 & 0.855 & 0.867 \\
\multirow{2}{*}{ESM2} & t30 & 153M & \usym{2713} & \usym{2713} & \usym{2717} & 91.8 & 87.4 & 86.9 & 88.9 & 98.4 & 87.4 & 0.923 & 0.874 & 0.843 & 0.837 \\
 & t33 & 655M & \usym{2713} & \usym{2713} & \usym{2717} & \textbf{\textcolor{red}{94.4}} & \textbf{\textcolor{third_score}{91.9}} & \textbf{\textcolor{third_score}{92.6}} & \textbf{\textcolor{red}{93.0}} & 96.4 & \textbf{\textcolor{third_score}{91.9}} & \textbf{\textcolor{red}{0.945}} & \textbf{\textcolor{third_score}{0.919}} & \textbf{\textcolor{red}{0.889}} & \textbf{\textcolor{red}{0.896}} \\
\multirow{2}{*}{ProtBert} & t30 Uniref & 423M & \usym{2713} & \usym{2713} & \usym{2717} & 92.8 & 90.3 & \textbf{98.9} & 89.6 & 86.6 & 88.2 & 0.923 & 0.902 & 0.863 & 0.872 \\
 & t30 BFD & 423M & \usym{2713} & \usym{2713} & \usym{2717} & 93.2 & 91.5 & 88.6 & 91.7 & \textbf{\textcolor{third_score}{99.2}} & 91.6 & 0.936 & 0.916 & 0.870 & 0.888 \\
 Ankh & base & 453M & \usym{2713} & \usym{2713} & \usym{2717} & 92.3 & 91.4 & 86.8 & 91.4 & \textbf{99.8} & 91.4 & 0.928 & 0.913 & 0.856 & 0.887 \\
\midrule
\multirow{2}{*}{SaProt} & t12 AF2 & 35M & \usym{2713} & \usym{2717} & \usym{2713} & 71.0 & 44.4 & 68.7 & 47.1 & 77.2 & 44.4 & 0.795 & 0.450 & 0.423 & 0.262 \\
 & t33 PDB & 650M & \usym{2713} & \usym{2717} & \usym{2713} & 75.8 & 46.7 & 73.0 & 46.5 & 82.0 & 46.6 & 0.727 & 0.459 & 0.520 & 0.292\\
 \midrule
 CPE-Pro & t3 Swiss-Prot & 29M & \usym{2717} & \usym{2713} & \usym{2713} & \textbf{98.5} & \textbf{97.2} & \textbf{\textcolor{red}{97.6}} & \textbf{97.3} & \textbf{\textcolor{red}{99.4}} & \textbf{97.2} & \textbf{0.985} & \textbf{0.972} & \textbf{0.970} & \textbf{0.963} \\
\botrule
\end{tabular}
\footnotetext{Note: AA: Amino Acids, struct-seq: ``structure-sequence''. All results are rounded to three decimal places and the best results are in bold. The top three are highlighted by \textbf{First}, \textbf{\textcolor{red}{Second}}, \textbf{\textcolor{third_score}{Third}}.}
\end{sidewaystable}

\section{Results and Analysis}\label{sec5}
\textbf{CPE-Pro demonstrates exceptionally high accuracy performance in structure discrimination tasks.} Baseline models and CPE-Pro were first trained on the C-A task. After several iterations, CPE-Pro achieved an accuracy of 0.98 on the test set, with the other four metrics all exceeding 0.97. In contrast, although the hybrid approach combining senven PLMs with GVP-GNN also achieved an accuracy above 0.9 in the C-A task, it still fell short compared to CPE-Pro. Among these senven hybrid baseline methods, the middle-performing model was ESM-1v, while the other six PLMs all had top-three performances across various metrics. The best accuracy was achieved by the method using a 33-layer ESM2 as the sequence encoder, reaching over 0.94—still  4.1\% lower than CPE-Pro. The worst-performing model was the 153M parameter ESM2, which, due to its smaller model size, had lower performance compared to the other six PLM methods. This suggests that the size and complexity of PLMs influence their ability to capture the structure and function of proteins to a certain extent, with smaller models possibly struggling to fully learn the deep connections between sequence and structure in complex tasks.

Building on this success, we extended training to the more complex C-M task. After training on multi-class structured data, the model's performance on the C-M task gradually improved, with all five metrics converging around 0.97, demonstrating strong competitiveness. However, both versions of SaProt performed poorly in the experiments, achieving less than 80\% accuracy in distinguishing between crystal and AlphaFold2-predicted structures, and failing to reach 50\% accuracy in the C-M task. A detailed analysis of the underlying reasons for this will be discussed in the subsequent section.

\begin{figure}[h]
\centering
\includegraphics[width=0.9\textwidth]{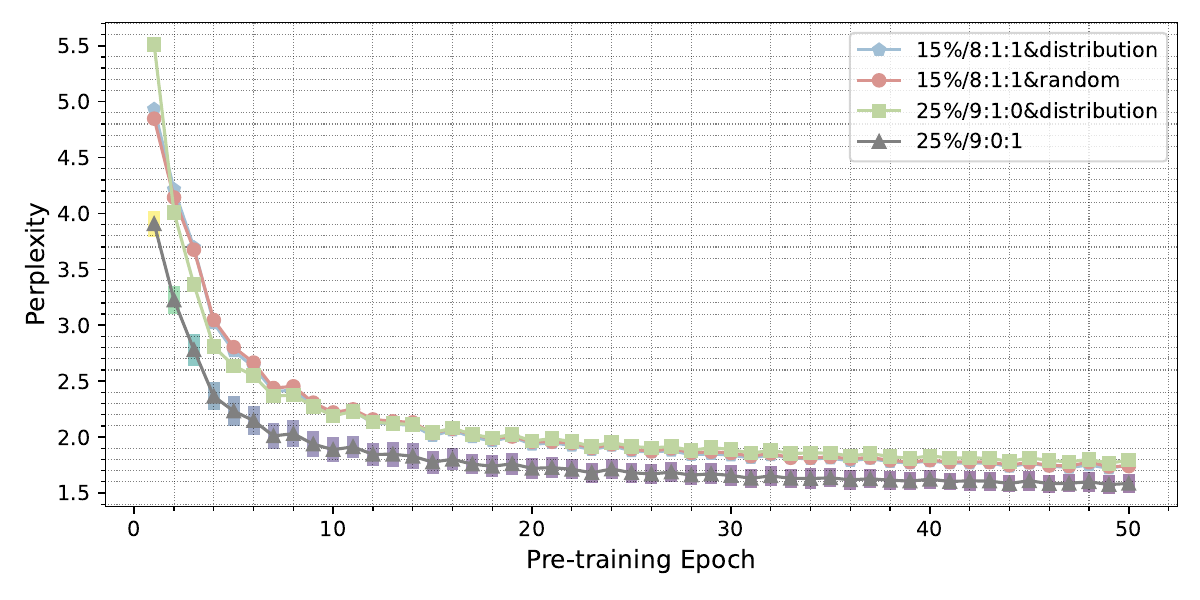}
\caption{The perplexity $ \downarrow $
 of SSLM on the validation set. Among the 4 training strategies, the combination of a 25\% masking rate with the 9:0:1 masking method shows superior performance. The original curve depicts how perplexity changes with training steps, while the smoothed curve illustrates its trend, reducing noise and providing a clearer view of the decreasing perplexity trend.}\label{pretrain_perplexity}
\end{figure}

\noindent \textbf{The ``structure-sequence'' can be a better predictor.} 
The pre-training of the protein structural sequence language model, SSLM utilized 109,334 high pLDDT score  (\textgreater 0.955) protein structures from the Swiss-Prot database. The masking strategy and rate used in the pre-training task were informed by the approach proposed in \cite{wettig2022should}, which comprehensively considers the model size and dataset scale. The setup of pre-training task are detailed in Table~\ref{pretrain_tab} and Figure~\ref{pretrain_perplexity} illustrates the perplexity of the SSLM on the pre-training task.

\begin{figure}[h]
\centering
\includegraphics[width=0.9\textwidth]{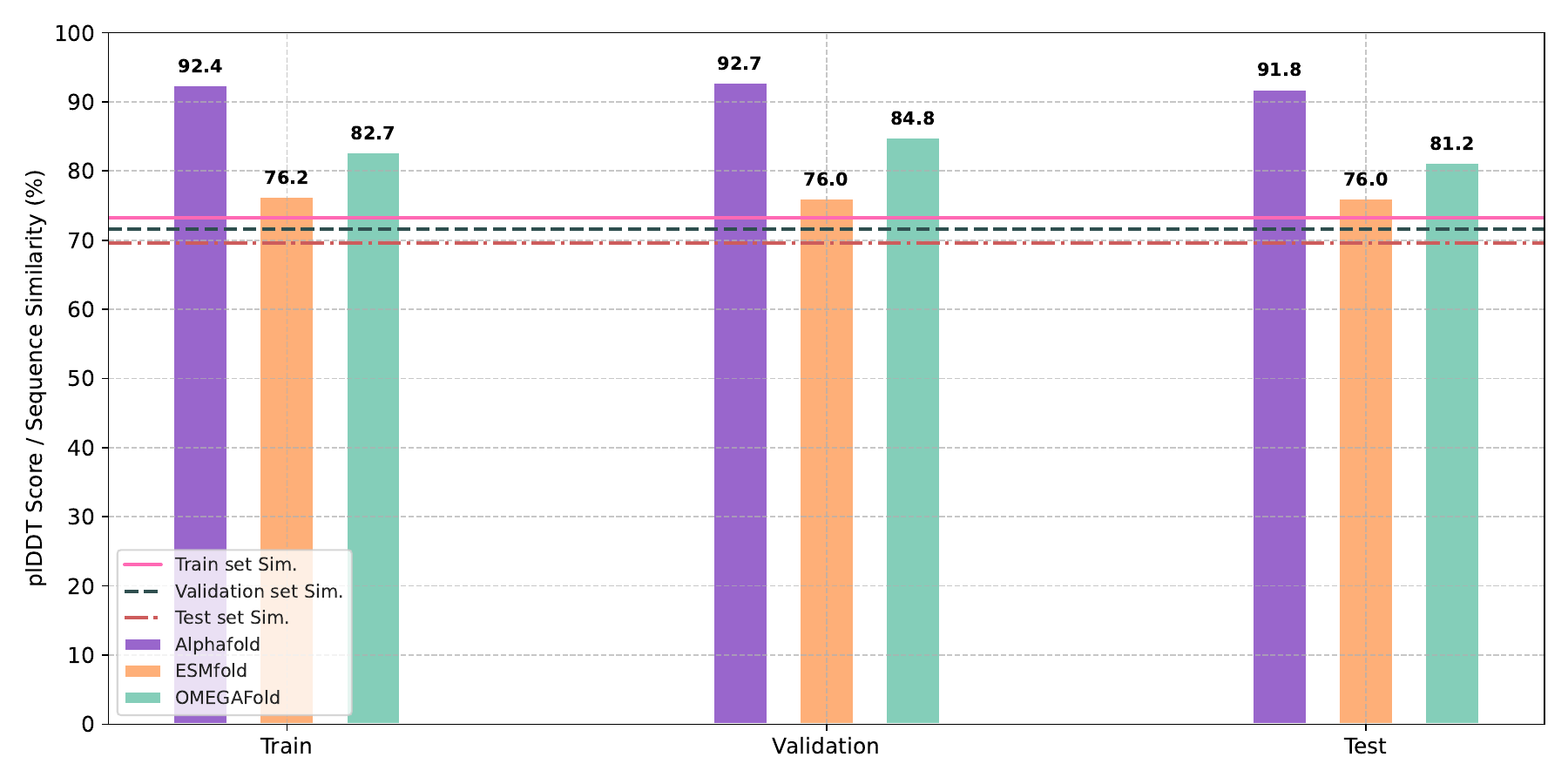}
\caption{The pLDDT scores of the predicted protein structures in the dataset used for training CPE-Pro and the similarity between ``structure-sequences''.}\label{plddt_and_sim}
\end{figure}

Sequential encoder SSLM in CPE-Pro and baseline models ESM1b, ESM1v, ESM2, ProtBert, Ankh were evaluated on the performance of CATH-PFD. Pre-trained SSLM has 3 hidden layers and significantly reduced parameters. The results in Table \ref{main_result} and \ref{ablation} show that the inclusion of ``structure-sequence'' encoders outperforms these amino acid sequence encoders on downstream C-M tasks. We preliminarily confirm our hypothesis on a structure discrimination task that language models learn sequence information obtained directly from structure discretization more efficiently. The ``structure-sequence'' shows greater effectiveness in protein classification tasks, which provides new directions for further optimization and design of more efficient predictive models. Independent use of SSLM and the structure-aware language model SaProt performed poorly because both rely solely on sequence input. Figure \ref{plddt_and_sim} shows the average pLDDT scores and similarity between ``structure-sequences'' for protein structures used in training, validation, and testing. It is evident that the high pLDDT levels across the three categories resulted in higher similarity of ``structure-sequences'' (training set: 73.28\%; validation set: 71.58\%; test set: 69.60\%), leading to homogenized feature representations and limited model generalization. In contrast, SaProt's input comes from a vocabulary of size 441 that includes both amino acid and ``structure-sequence'' elements. By combining these two vocabularies, some effects of high similarity are mitigated, and its performance metrics improve compared to SSLM.
To demonstrate the effectiveness of SSLM in capturing structural differences, we validated it through visualization methods in subsequent sections. We also speculate that the scaling effect applies to language models trained with ``structure-sequences''. In other words, increasing the model's depth and the scale of training data could significantly enhance the model’s performance. Even without relying on a structural encoder, the model may still achieve satisfactory results.

\begin{table}[h]
\caption{Pre-training Settings of SSLM}\label{pretrain_tab}%
\begin{tabular}{lccc}
\toprule
Model & Data Size & Mask Radio & Mask Method \\ 
\midrule
\multirow{4}{*}{SSLM\_t3\_25M\_Swiss-Prot} & \multirow{4}{*}{109,334} & \multirow{2}{*}{15\%} & 8:1:1\&random \\
 & & & 8:1:1\&distribution \\
 & & \multirow{2}{*}{\textbf{25\%}} & \textbf{9:0:1} \\
 & & & 9:1:0\&distribution \\
\botrule
\end{tabular}
\footnotetext{The pre-training process employed four masking strategies. For the SSLM\_t3\_25M\_CATH-PFD model, with a 15\% masking rate, two masking methods were used: an 8:1:1 ratio with random replacement and an 8:1:1 ratio with distribution. The distribution reflects the statistical number of tokens in the pre-training dataset. When the masking rate increased to 25\%, the masking strategies were adjusted to 9:0:1 and 9:1:0 with distribution. These configurations were designed to explore the performance of the model pre-trained using the MLM task.}
\end{table}

\begin{figure}[h]
\centering
\includegraphics[width=0.9\textwidth]{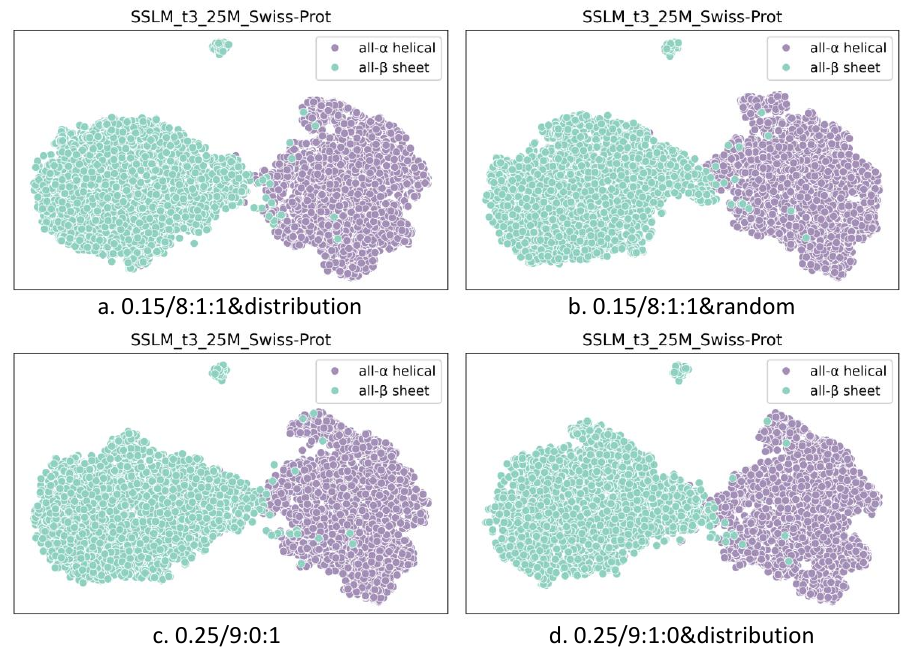}
\caption{Using the t-SNE method, the feature embeddings of four pre-trained versions of SSLM in the SCOPe database were dimensionally reduced and visualized on a two-dimensional plane.}\label{t-sne_sslm}
\end{figure}

\begin{figure}[h]
\centering
\includegraphics[width=0.9\textwidth]{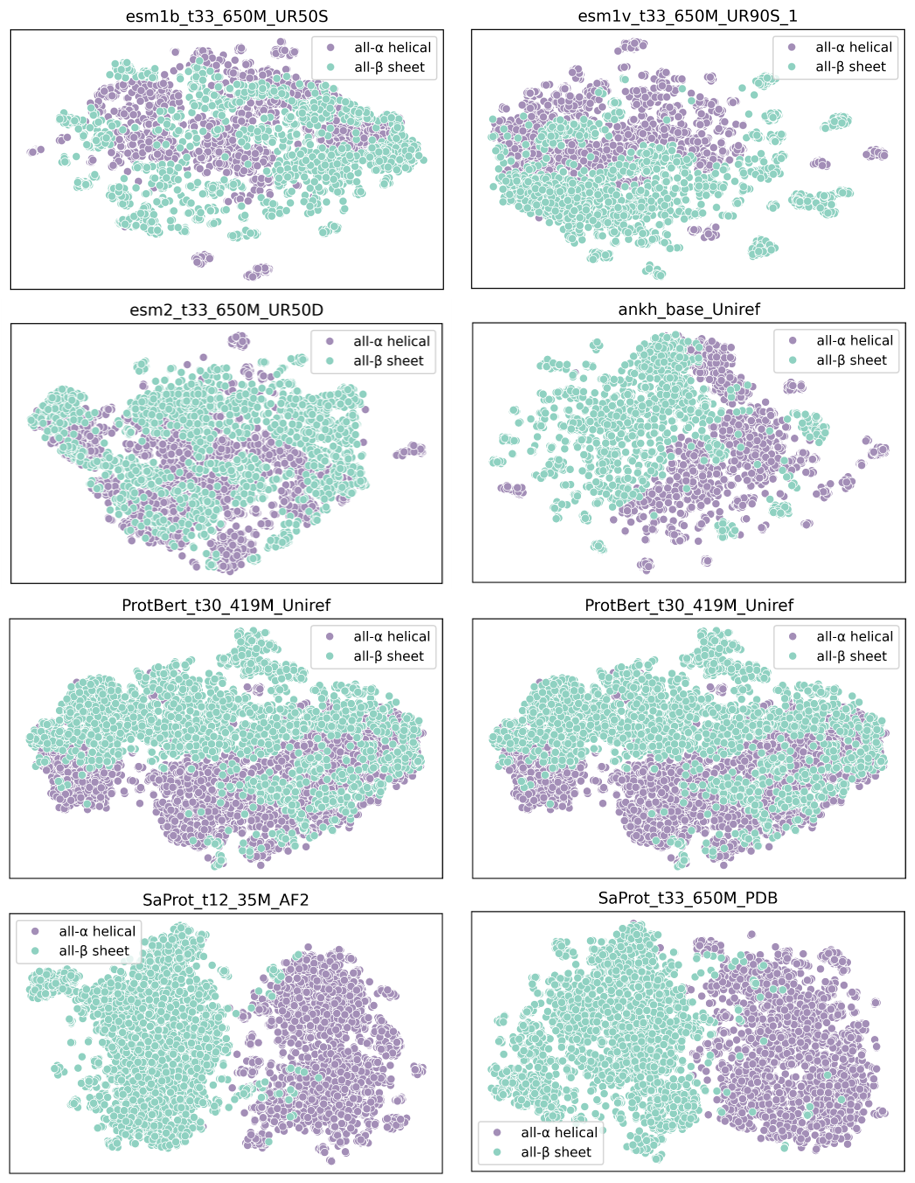}
\caption{Using the t-SNE method, the feature embeddings of various PLMs in the SCOPe database were dimensionally reduced and visualized in a two-dimensional plane.}\label{t-sne_plm}
\end{figure}

\begin{table}[h]
\caption{Ablation Study Results on CPE-Pro}\label{ablation}%
\begin{tabular}{cccccccccc}
\toprule
\multirow{2}{*}{GVP-GNN} & \multicolumn{2}{c}{SSLM} & \multirow{2}{*}{\shortstack{Attention \\ Pooling}} & \multicolumn{2}{c}{Acc(\%) $ \uparrow $
} & \multicolumn{2}{c}{F1-score $ \uparrow $
} & \multicolumn{2}{c}{MCC $ \uparrow $
} \\
\cmidrule(lr){2-3} \cmidrule(lr){5-6} \cmidrule(lr){7-8} \cmidrule(lr){9-10}
 & P-t & W/o P-t & & C-A & C-M & C-A & C-M & C-A & C-M \\
\midrule
\usym{2717} & $\usym{2713}$ & \textbf{--} & \usym{2713} & 68.5 & 39.8 & 0.694 & 0.372 & 0.371 & 0.204 \\
\usym{2713} & \usym{2717} & \textbf{--} & \usym{2713} & 90.4 & 87.2 & 0.912 & 0.867 & 0.823 & 0.839 \\
\usym{2713} & \usym{2713} & \textbf{--} & \usym{2717} & 95.9 & 94.9 & 0.961 & 0.949 & 0.921 & 0.933 \\
\usym{2713} & \textbf{--} & \usym{2713} & \usym{2713} & 98.1 & 92.2 & 0.981 & 0.922 & 0.962 & 0.897 \\
\usym{2713} & \usym{2713} & \textbf{--} & \usym{2713} & \textbf{98.5} & \textbf{97.2} & \textbf{0.985} & \textbf{0.972} & \textbf{0.970} & \textbf{0.963} \\
\botrule
\end{tabular}
\footnotetext{P-t: Pre-trained; Attention Pooling: pooling layer with attention mask. The experiments vary by removing components like SSLM, GVP-GNN, and AM-Pooling to test their impact. Note: The best results are in bold.}
\end{table}

\begin{figure}[h]
\centering
\includegraphics[width=0.9\textwidth]{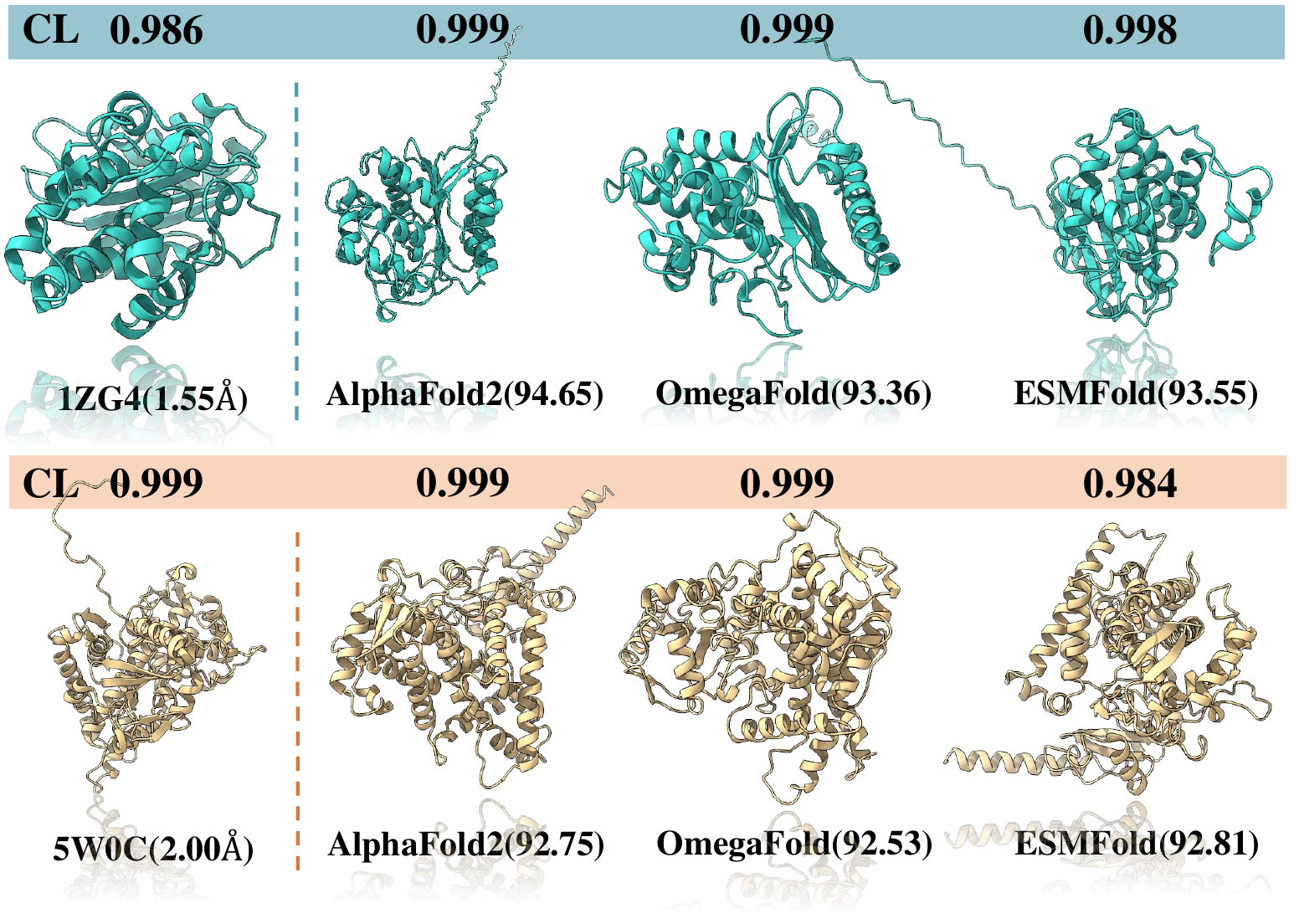}
\caption{The X-ray crystal structures 1BT5 and 5W0C , along with the outputs from three predictive models for each structure, were selected for the case study. All predicted structures achieved pLDDT scores above 90. The confidence levels (CL) demonstrate the high accuracy and robustness of CPE-Pro in structural origin evaluation tasks.}\label{case-study}
\end{figure}

\noindent \textbf{Feature Visualization Method Powerfully Demonstrates Pretrained SSLM's Excellence in Capturing Structural Differences.} The protein language model has been shown to embed secondary and tertiary structure\cite{rives2021biological} characteristics within its output representations of proteins. We selected a subset of gene domain sequences from the non-redundant Astral SCOPe 2.08 database in SCOPe\cite{chandonia2019scope}, where the identity between sequences is less than 40\%. From this subset, we focused on all-\( \alpha \) helical proteins (2,644) and all-\( \beta \) sheet proteins (3,059) and filtered the corresponding structural sets in the database. Figure~\ref{t-sne_sslm} and~\ref{t-sne_plm} show the t-SNE visualization of the protein representations from the last hidden layers of SSLM and various PLMs on the aforementioned dataset. It is evident that, aside from SaProt, which incorporates ``structure-sequence'' in its input, the representations of other PLMs, while capturing some differences in structural types, exhibit relatively weak discriminative power. After dimensionality reduction, the distribution of data points becomes chaotic, and the boundaries between the protein classes are blurred. In contrast, both SaProt and SSLM not only differentiate the two protein classes more effectively but also provide clearer class boundaries, with SSLM showing a more concentrated distribution within each class. This suggests that SSLM possesses stronger discriminative capability in capturing and representing protein structural features, providing a more accurate reflection of the intrinsic characteristics of different structural types.

\noindent \textbf{Ablation Study on Components of CPE-Pro.} To validate the contribution of each component designed in CPE-Pro, we conducted five sets of ablation experiments on both the C-A and C-M tasks. These variations included removing the GVP-GNN, omitting the pre-training process of SSLM, removing the SSLM, eliminating attention in the pooling layer, and using all three components together. As shown in Table~\ref{ablation}, each component made a positive contribution to the C-M task. Performance significantly declined when using only a single type of encoder, particularly when using the SSLM alone as discussed earlier. The CPE-Pro model, which employs a pre-trained SSLM, outperformed the non-pre-trained version, indicating that the pre-training process effectively helped the model learn the structural features embedded in ``structure-sequences''. Additionally, the application of attention-masked pooling layers also positively influenced model performance, further enhancing overall effectiveness.

\noindent \textbf{Case Study: Discrimination of the Structural Origins of BLAT ECOLX and CP2C9 HUMAN.} BLAT ECOLX is a \( \beta \)-lactamase protein found in Escherichia coli that can hydrolyze the \( \beta \)-lactam ring of \( \beta \)-lactam antibiotics, rendering these antibiotics inactive. It plays a significant role in the study of antibiotic resistance. CP2C9 HUMAN is a human Cytochrome P450 2C9 enzyme responsible for the metabolism of various drugs, including non-steroidal anti-inflammatory drugs and anticoagulants. It plays a significant role in drug metabolism and the regulation of endogenous substances. In three structural prediction models, both proteins achieved pLDDT scores above 0.9, indicating high accuracy in structure prediction with minimal deviation from the crystal structure. We input the crystal structures and predicted structures of both proteins into CPE-Pro for origin evaluation. Figure~\ref{case-study} demonstrates that the model successfully and confidently predicted the origins of these structures. This result highlights the robustness of the model in assessing structural origins, even in cases of very minor structural differences.

\section{Conclusion}\label{sec6}

In this study, we developed a protein folding dataset, CATH-PFD, derived from the non-redundant dataset in the CATH database, which incorporates structures from various prediction models. By training and validating our model, CPE-Pro, on the CATH-PFD dataset, we created an innovative and effective solution for identifying the structural origins of proteins. The CPE-Pro model excels in learning and analyzing protein structural features, outperforming methods that combine amino acid sequences with structural data, as well as models that use structure-aware sequences, in the task of structural origin recognition. In case studies, CPE-Pro demonstrated superior performance.

These findings provide preliminary evidence that incorporating ``structure-sequence'' information significantly enhances the language model's ability to learn protein features, enabling it to capture richer and more precise structural details, thereby improving the representation of protein structures. In subsequent visualization experiments, we further validated the sensitivity of SSLM, utilized within CPE-Pro, to structural variations, as well as its effectiveness in capturing and representing complex protein structural features. This exploration not only opens up new avenues for the application of protein structural language models but also paves the way for future developments in the paradigm and interpretability of protein structure prediction methods, offering fresh insights and possibilities for advancing practical applications in bioinformatics and structural biology.
\backmatter

\section*{Declarations}

\begin{itemize}
\item \textbf{Competing interests} The authors declare that they have no known competing financial interests or personal relationships that could have appeared to influence the work reported in this paper.  
\end{itemize}

\bibliography{sn-bibliography}

\end{document}